\documentclass[twocolumn,aps,amsmath,amssymb]{revtex4}
\usepackage{graphicx}
\usepackage{dcolumn}
\usepackage{bm}

\def\phc{\bar}
\def\subt#1{\mathrm{#1}}  

\def\mystar{*}

\begin{document}

\preprint{?}
\title{Composite Fermions in Negative Effective Magnetic Field:  A Monte-Carlo Study}

\author{Gunnar M\"oller${}^{a,b}$ and Steven H. Simon${}^a$}
\affiliation{ ${}^a$Bell Laboratories, Lucent Technologies,
Murray Hill, New Jersey 07974
\\${}^b$Laboratoire de Physique Th\'eorique et M\'ecanique Statistique, 91406 Orsay, France
}

\date{February 21, 2005}

\begin{abstract}
The method of Jain and Kamilla [PRB {\bf 55}, R4895 (1997)] allows
numerical generation of composite fermion trial wavefunctions for
large numbers of electrons in high magnetic fields at filling
fractions of the form $\nu=p/(2mp+1)$ with $m$ and $p$ positive
integers. In the current paper we generalize this method to the
case where the composite fermions are in an effective (mean) field
with opposite sign from the actual physical field, i.e. when $p$
is negative. We examine both the ground state energies and the low
energy neutral excitation spectra of these states. Using
particle-hole symmetry we can confirm the correctness of our
method by comparing results for the series $m=1$ with $p>0$
(previously calculated by others) to our results for the conjugate
series $m=1$ with $p <0$. Finally, we present similar results for
ground state energies and low energy neutral excitations for the
states with $m=2$ and $p <0$ which were not previously
addressable, comparing our results to the $m=1$ case and the $p >
0$, $m=2$ cases.
\end{abstract}
\maketitle

\section{Introduction}

The composite fermion approach\cite{Olle} has had a great number
of extremely impressive successes in describing the physics of
electrons in high magnetic fields.  In this picture, fractional
quantum Hall systems in total magnetic field $B$ are described in
terms of noninteracting ``composite fermions" in an effective
magnetic field $B_{\subt{eff}} = B - 2 m \phi_0 n$, where $\phi_0
= hc/e$ is the flux quantum, $n$ is the electron density, and $m$
is a positive integer.  This maps, for example, fractional quantum
Hall states at filling fractions of the form $\nu=n \phi_0/B =
p/(2mp+1)$ to integer quantum Hall states for the composite
fermions at filling fraction $\nu_{\subt{eff}} = n
\phi_0/|B_{\subt{eff}}| = |p|$.  We will denote such composite
fermions with $2m$ flux quanta attached to them as $^{2m}\!$CF.

Jain's original approach to composite fermions\cite{Jainbook}
constructed highly accurate trial wavefunctions by taking simple
wavefunctions for the noninteracting (composite) fermions in the
effective magnetic field, multiplying by Jastrow factors,
and then projecting the result into the lowest Landau level. The
early successes of this method were impressive\cite{Jainbook},
despite the fact that the method was limited by the extreme
numerical difficulty of performing projections for systems with
more than roughly 10 electrons.

A major theoretical breakthrough came when Jain and
Kamilla\cite{JainKamilla} discovered a new way of writing
composite fermion trial wavefunctions (described below), which
involves a very minor modification of the projection. These new
trial states seemed to be just as good as the originally proposed
wavefunctions and could be numerically generated even for systems
with many electrons (40 electrons or more).  Since that time, many
important studies have been achieved using this
method\cite{JainKamillaExciton,JainMass,MoreJainMass,OtherJainPapers}.
However, so far this method has been restricted to cases where the
effective magnetic field has the same sign as the external
magnetic field.  Results using this method have been published for
filling fractions of the form $\nu=p/(2mp+1)$ with $p > 0$ but not
for $p <0$.   In the current paper, we extend the work of Jain and
Kamilla\cite{JainKamilla} so that we are able to handle states
with $p<0$.  The $p < 0$ states take more computational resources
than the case of $p>0$, and the difference in the computational
resources between the two cases increases with the absolute
value of the effective flux.  However, the computational problems
turn out to be more severe for the smallest $|p|$, where the number
of particles in the system increases most slowly with each flux
added. Fortunately, we probably need not go to particularly large
systems to understand the physics of small $|p|$. For large $|p|$,
describing the approach to the Fermi liquid-like composite fermion
state, the system size is already large for relatively small
effective flux, i.e. the differences of the computational
requirements in the case of $p>0$ and $p<0$ become relatively less
important.

Although the case of negative $p$ has not previously been studied
for large systems, we point out that for $m=1$ the series of
states with negative $p$ and the series of states with positive
$p$ are essentially equivalent due to an exact particle-hole
symmetry in the lowest Landau level. In fact, below, we exploit
this symmetry to check the validity of our method. Once we have
verified the method, we can study the properties of the $m=2$
series for negative $p$ and compare the results to those of the
positive $p$ members of this same series as well as to those of
the $m=1$ series.

The outline of this paper is as follows.  In section \ref{sec:jk}
we briefly review the Jain-Kamilla method.  As mentioned above,
the method has only been used previously for the case of positive
$p$.  In appendix \ref{app:jk} we show in detail how the crucial
projection scheme of Jain-Kamilla can easily be generalized to
handle negative $p$ also. It is easy to see from the result how
much additional numerical complexity is involved for negative $p$.
In section \ref{sec:phsym} we test our approach by examining the
 ${}^2$CF series $\nu=p/(2p+1)$.  In particular, we examine ground state
energies, excitation spectra, and energy gaps.  We pay particular
attention to the mass of the composite fermion and the scaling of
the gap with $p$.   For $p>0$, results are already available in
the
literature\cite{JainKamilla,JainKamillaExciton,JainMass,MoreJainMass,OtherJainPapers}.
For $p<0$ we use our generalization of the Jain-Kamilla method to
calculate energies directly, and we compare these energies to
energies obtained by particle-hole conjugating the $p>0$ series.
Appendix \ref{app:phconj} describes the particle-hole conjugation
transformation in depth.  This comparison establishes the accuracy
of our method.   In section \ref{sec:cf4results} we move on to
examine the ${}^4$CF series of states $\nu=p/(4p + 1)$.    Again,
for $p>0$ some results are already available in the
literature\cite{JainKamilla,JainKamillaExciton,JainMass,MoreJainMass,OtherJainPapers}.
However, for $p<0$ our results are new. Again, we examine ground
state energies, excitation spectra, energy gaps, and composite
fermion effective masses. (By using particle-hole conjugation, we
could give results for filling fractions $\nu=1 -p/(4p + 1)$)
similarly). We are able to make some comparison of our energy gaps
to the experimental work of Pan \emph{et al.}\cite{Pan}.

Throughout this paper we assume complete spin-polarization of the
electrons.  This should be a reasonable assumption for real
experiments at sufficiently high magnetic fields.

\section{The Jain-Kamilla Method}
\label{sec:jk}

Jain's original proposal\cite{Jainbook} was to construct trial
wavefunctions for fractional quantum Hall states by writing
\begin{equation}
\label{eq:jain0}
   \Psi_{\subt{Jain}} = {\cal P} \left\{ \det\left[ \psi_i(\vec r_j) \right] \Phi_0^{2m} \right\}
   \end{equation}
where the determinant is a Slater determinant of noninteracting
single fermion wavefunctions $\psi_i$ in effective magnetic field
$B_{\subt{eff}} = B - 2 m n \phi_0$, and ${\cal P}$
indicates projection to the lowest Landau level.  Here,
$\Phi_0$ is the wavefunction of a completely filled Landau level,
\begin{equation}
   \Phi_0 = \prod_{i < j} (z_i - z_j)
\end{equation}
where $z_j = (x_j + i y_j)/ \ell_0$ is the dimensionless complex
coordinate on the plane, $\ell_0=\sqrt{\hbar c/eB}$ and the usual Gaussian factors
$\exp(-\frac{1}{4}\sum |z_i|^2)$ are understood to be included in
the measure of the Hilbert space and will not be written
explicitly for simplicity of notation.

Choosing a set of single particle wavefunctions $\psi_i$ to fill
the $p$ lowest effective Landau levels (i.e., such that
$\det[\psi_i(\vec r_j)]$ represents the ground state of an {\it
integer} quantum Hall state $\nu=p$), one obtains through Eq.
\ref{eq:jain0} extremely good trial wavefunctions for fractional
quantum Hall states $\nu=p/(2mp+1)$.   As discussed above, the
projection in Eq. \ref{eq:jain0} is exceedingly hard to implement
for
systems with more than roughly 10 electrons.  For this reason,
Jain and Kamilla\cite{JainKamilla} looked for an essentially
equivalent formulation that would be computationally simpler.
In their approach they begin by rewriting the wavefunction as
\begin{equation}
 \Psi_{\subt{Jain}} = {\cal P} \left\{ \det\left[ \psi_i(\vec r_j)  J_j^m \right]  \right\}
\end{equation}
where
\begin{equation}
J_j = \prod_{k \neq j}
  (z_k - z_j)
\end{equation}
and then make the approximation that one can interchange the order of
projection and taking the determinant to obtain a new trial
wavefunction
\begin{equation}
 \Psi_{\subt{JK}} =   \det\left[ \tilde \psi_i(\vec r_j) \right]
 \Phi_0^{2m}
\end{equation}
with
\begin{equation}
\label{eq:tildepsi1} \tilde \psi_i(\vec r_j) = J_j^{-m} {\cal P} \left\{
\psi_i(\vec r_j) J_j^m \right\}
\end{equation}
Although $\Psi_{\subt{JK}}$ appears to be a single Slater
determinant, it is somewhat more complicated because each $\tilde
\psi_i(\vec r_j)$ is actually a function of all of the particle
positions through $J_j$. Nonetheless, this new trial wavefunction
is far simpler to evaluate numerically. Furthermore, extensive
numerical
work\cite{JainKamilla,JainKamillaExciton,JainMass,MoreJainMass,OtherJainPapers}
has shown that for small systems $\Psi_{\subt{JK}}$ is just as
good a trial state as $\Psi_{\subt{Jain}}$ and that both are
extremely accurate\footnote{We note in passing, that one could
certainly construct several other trial wavefunctions which
project in slightly different ways. For example, for $m>1$ we
might consider $ \tilde \psi_i(\vec r_j) = {\cal P} \left\{
\psi_i(\vec r_j) J_j^{m_1} \right\}  J_j^{m_2} $ with $m_1 + m_2 =
m$ (and $m_1 \geq 1$). In the cases we have checked we have found
that wavefunctions built with this version of $\tilde \psi$ are
also extremely similar to those built from Eq.
\ref{eq:tildepsi1}.}. In the original work by Jain and Kamilla, it
was shown how to calculate $\tilde \psi$ on a sphere for the case
when $B_{\subt{eff}}$ has the same sign as the magnetic field $B$
(i.e., $p > 0$). In Appendix \ref{app:jk} we repeat the derivation
for the case where $B_{\subt{eff}}$ has the opposite sign from $B$
(i.e., $p <0$).  A discussion is also given there of the relative
computational effort required to perform the relevant computations
numerically.

This technique allows one to also obtain low energy spectra of
these fractional quantum Hall states, by similarly
composite-fermionizing low energy excited states of noninteracting
fermions as discussed in Ref. \onlinecite{JainKamillaExciton}.

In this paper we will perform all calculations using a spherical
geometry\cite{Haldane} with a monopole of charge $N_{\phi}$ flux
quanta at the center.  The composite fermions then see an
effective flux $2q = N_{\phi}^{\subt{eff}} = N_{\phi} - 2m (N-1)$.
In the presence of this effective flux, single particle states are
described by two quantum numbers, $l$ and $m$. Here $l = |q|+n$ is
the angular momentum with $n=0,1,2,\ldots$ corresponding to the
``Landau level number" or ``shell" index, and $m$ is the
$z$-component of angular momentum.  A state with $p$ filled
composite fermion Landau levels corresponds to $n=|p|-1$.

A low energy exciton is now formed by taking a composite fermion
out of the highest occupied  shell  (or Landau level) $l = l_F =
|q| + |p|-1$ with some $m_h$ and putting it in the lowest
unoccupied $l= l_{F} + 1$ shell with some $m_e$.  Choosing to work
with states of zero total $z$-angular momentum,  we take the state
with $m_e = -m_h$ and write this state as $| m_e \rangle$. Using
vector coupling (Clebsh-Gordon) coefficients\cite{Edmonds} we can
construct exciton eigenstates of angular momentum $L$
as\cite{JainKamillaExciton}
\begin{equation}\label{eq:excitonadd}
 \Xi_L^{\text{exciton}} = \sum_{m_e=-l_F}^{l_F}|m_e\rangle \langle l_F,-m_e;l_F+1,m_e|L,0\rangle
\end{equation}
which serve as extremely accurate trial wavefunctions for the low
energy excited states of the above discussed composite fermion
ground states.

\begin{table*}[t!]

\begin{tabular}{||c||c|c|c|D{.}{.}{6.8}|D{.}{.}{4.7}|D{.}{.}{4.7}||c|c|c|D{.}{.}{6.9}|D{.}{.}{4.7}|D{.}{.}{3.8}||}
\colrule \multicolumn{1}{||c||}{} & \multicolumn{6}{c||}{~~~~~~~~
Negative $p$ Trial Wavefunction~~~~~~~~~} &
\multicolumn{6}{c||}{~~~~~ Positive $p$ Trial Wavefunction~~~~~~}\\
\colrule $N_{\phi}$& $p$&$N_{\phi}^{\text{eff}}$&$\; N\;$
&\multicolumn{1}{c|}{$E_g$}&\multicolumn{1}{c|}{MR-gap}&
\multicolumn{1}{c||} {large k gap}& $\, p\,$ &$N_{\phi}^{\text{eff}}$&$\; N\;$  &
\multicolumn{1}{c|}{ $E_{\bar g}$ : P-H conj}&\multicolumn{1}{c|}
{MR-gap}&\multicolumn{1}{c||}{large k gap} \\
\toprule
9&-2&-1&6&-0.5391(1)&0.0906(9)&0.118(1)&1&3&4&-0.53949(2)&0.0928(2)&0.118(2)\\
12&&-2&8&-0.5338(1)&0.091(1)&0.115(1)&&4&5&-0.53412(2)&0.0931(2)&0.1124(4)\\
15&&-3&10&-0.5303(1)&0.083(2)&0.109(2)&&5&6&-0.53090(2)&0.0838(3)&0.1079(4)\\
18&&-4&12&-0.5282(1)&0.080(2)&0.104(3)&&6&7&-0.52873(2)&0.0801(4)&0.1040(5)\\
21&&-5&14&-0.5266(1)&0.078(2)&0.101(2)&&7&8&-0.52721(2)&0.0824(4)&0.1021(4)\\
24&&-6&16&-0.5257(1)&0.079(5)&0.100(3)&&8&9&-0.52607(2)&0.0768(5)&0.1013(6)\\
\cline{1-1}\cline{3-7} \cline{9-13}
$\infty$&&-$\infty$&$\infty$&-0.5173(1)&0.070(3)&0.089(3)&&$\infty$&$\infty$&-0.51803(3)&0.069(4)&0.0907(9)\\
\toprule
21&-3&-1&12&-0.4985(1)&0.061(2)&0.067(2)&2&3&10&-0.49870(5)&0.061(1)&0.069(1)\\
26&&-2&15&-0.4980(1)&0.056(3)&0.063(3)&&4&12&-0.49826(5)&0.054(1)&0.066(1)\\
31&&-3&18&-0.4979(1)&0.061(3)&0.067(3)&&5&14&-0.49803(5)&0.053(1)&0.063(1)\\
36&&-4&21&-0.4979(1)&0.056(3)&0.069(3)&&6&16&-0.49804(5)&0.053(1)&0.065(1)\\
41&&-5&24&-0.4976(1)&0.048(4)&0.058(4)&&7&18&-0.49797(6)&0.050(2)&0.064(2)\\
\cline{1-1}\cline{3-7}\cline{9-13}
$\infty$&&-$\infty$&$\infty$&-0.4968(2)&0.042(8)&0.060(8)&&$\infty$&$\infty$&-0.4972(2)&0.041(3)&0.058(2)\\
\toprule
37&-4&-1&20&-0.4847(1)&0.052(3)&0.052(3)&3&3&18&-0.48467(5)&0.051(1)&0.053(2)\\
44&&-2&24&-0.4851(1)&0.050(3)&0.050(3)&&4&21&-0.48525(7)&0.041(2)&0.052(2)\\
51&&-3&28&-0.4855(1)&0.042(4)&0.048(4)&&5&24&-0.48564(8)&0.041(3)&0.048(3)\\
58&&-4&32&-0.4857(1)&0.036(4)&0.043(4)&&6&27&-0.48593(6)&0.041(3)&0.046(3)\\
\cline{1-1}\cline{3-7}\cline{9-13}
$\infty$&&-$\infty$&$\infty$&-0.4875(1)&0.014(7)&0.030(6)&&$\infty$&$\infty$&-0.48802(4)&0.022(8)&0.036(3)\\
\colrule
\end{tabular}

\caption{Numerical results for energies and gaps at filling
fractions $\nu=p/(2p+1)$, given in units of $e^2/\epsilon \ell_0$.
Calculations were performed using Monte-Carlo for ${}^2$CF trial
wavefunctions described in the text using $10^7$ samples.
Particle-hole conjugate pairs should give precisely the same
excitation energies (and the same ground state energies once Eq.
\ref{eq:pheq} is used, as it is here, see below). In other words,
if our results were exact, the right hand columns with positive
$p$ should precisely match the left hand columns with negative
$p$. Here, since we have used trial wavefunctions which are
approximate (albeit extremely good), the agreement is not quite
perfect, but it is extremely close. We note that the energies
obtained by using the positive $p$ states are slightly lower,
which means that the trial wavefunctions with positive $p$
(positive flux seen by the composite fermions) yield slightly
better trial states.   The number of electrons $N$ for
particle-hole conjugate pairs sums to $N_{\phi}+1$, which is one
filled Landau level. In this table, the ground state energy
$E_{g}$ is presented for the negative $p$ case.  For the positive
$p$ case, in the column labeled ``$E_{\bar g}$: P-H conj" we have
put the calculated ground state energy into Eq. \ref{eq:pheq} and
presented the result for comparison with the corresponding
negative $p$ states. The mageto-roton gap (MR-gap) is defined to
be the lowest energy neutral excitation. The large $k$ gap is the
gap measured at the highest possible angular momentum $L$ that we
can construct using Eq. \ref{eq:excitonadd} which is
$L_{\subt{max}} = 2l_F + 1$. All values indicated for the
thermodynamic limit have been extrapolated by a simple linear
regression over the inverse particle number using this set of data
only.} \label{tab:results_m1}
\end{table*}

\section{Particle-Hole Symmetry and Results for $\nu = p/(2p+1)$}
\label{sec:phsym}

Using particle-hole symmetry of the lowest Landau level, one can
exactly map states at filling fraction $\nu$ into states at
filling fraction $1 - \nu$ (so long as we maintain complete spin
polarization).  As mentioned above, in this paper we perform all
calculations using a spherical geometry\cite{Haldane}. On the
sphere, the lowest Landau level has $N_{\phi} + 1$ single particle
eigenstates with $N_{\phi}$ the total number of flux quanta
penetrating the sphere. Thus states with $N$ electrons can be
precisely mapped to their particle-hole conjugate states with
$N_{\phi} + 1 - N$ electrons (i.e. with $N$ holes). In Appendix
\ref{app:phconj} we show that, on a sphere, given an eigenstate
$\Psi$ with $N$ electrons and energy $E_\Psi$ one can write the
energy $E_{\phc \Psi}$ of its particle-hole conjugate
wavefunction $\phc \Psi$ as
\begin{equation}
\label{eq:pheq} E_{\phc \Psi}=\Bigl( 1-\frac{2 N}{N_\phi+1}
\Bigr) E_{\subt{filled}} + E_{\Psi}.
\end{equation}
where $E_{\subt{filled}}$ is the energy of the completely filled Landau
level.   This, of course, implies that the excitation spectrum of
any given state is precisely the same as the excitation spectrum
of its particle-hole conjugate state.

We will now focus on ${}^2$CF states of the form $\nu=p/(2p+1)$.
The state with $p$ is particle-hole conjugate of the state with $p
\rightarrow (-p -1)$.    (For example, $\nu=1/3$ which is $p=1$ is
conjugate of $2/3$ which is $p=-2$).   Extensive numerical work
has already been performed for positive $p$, calculating accurate
ground state energies and energy
gaps\cite{JainKamilla,JainKamillaExciton,JainMass,MoreJainMass,OtherJainPapers}.
Using Eq. \ref{eq:pheq} this means that we already know the ground
state energies and energy gaps for negative $p$. Here, however, we
calculate these quantities directly using our negative $p$ trial
wavefunctions and compare to the particle hole conjugated results
to establish the validity of our approach.

Table \ref{tab:results_m1} summarizes the numerical results for
the ground state energies of states in this series calculated
using
Monte-Carlo\cite{JainKamilla,JainKamillaExciton,JainMass,MoreJainMass,OtherJainPapers}
\footnote{A subtlety in this process results from the observation
that the wavefunction, calculated as described in the appendix
\ref{app:jk} by separating it into a Slater-determinant of the
projected pseudo single-particle-wavefunctions and a
Jastrow-factor, tends to produce numerical instabilities in the
determinant algorithm. This is easily understood, since two
particles approaching each other at a distance $d$ will cause the
respective $\tilde\Psi_i$ to grow as  $\tilde\Psi_i \propto
d^{-(|q|+n)}$ in the same manner, producing two linearly dependant
columns in the matrix. Now, due to numerical errors, instead of
obtaining zero when two columns become the same, the evaluation
yields numerical errors following approximately the same power law
behavior, that will dominate the result. Since the short distance
behavior of the wavefunction is known, extrapolation allows us to
correct for such events that occur with a probability of less than
$10^{-4}$ in all of the presented calculations.}.

\begin{figure}[bthp]
\includegraphics[width=0.95\columnwidth]{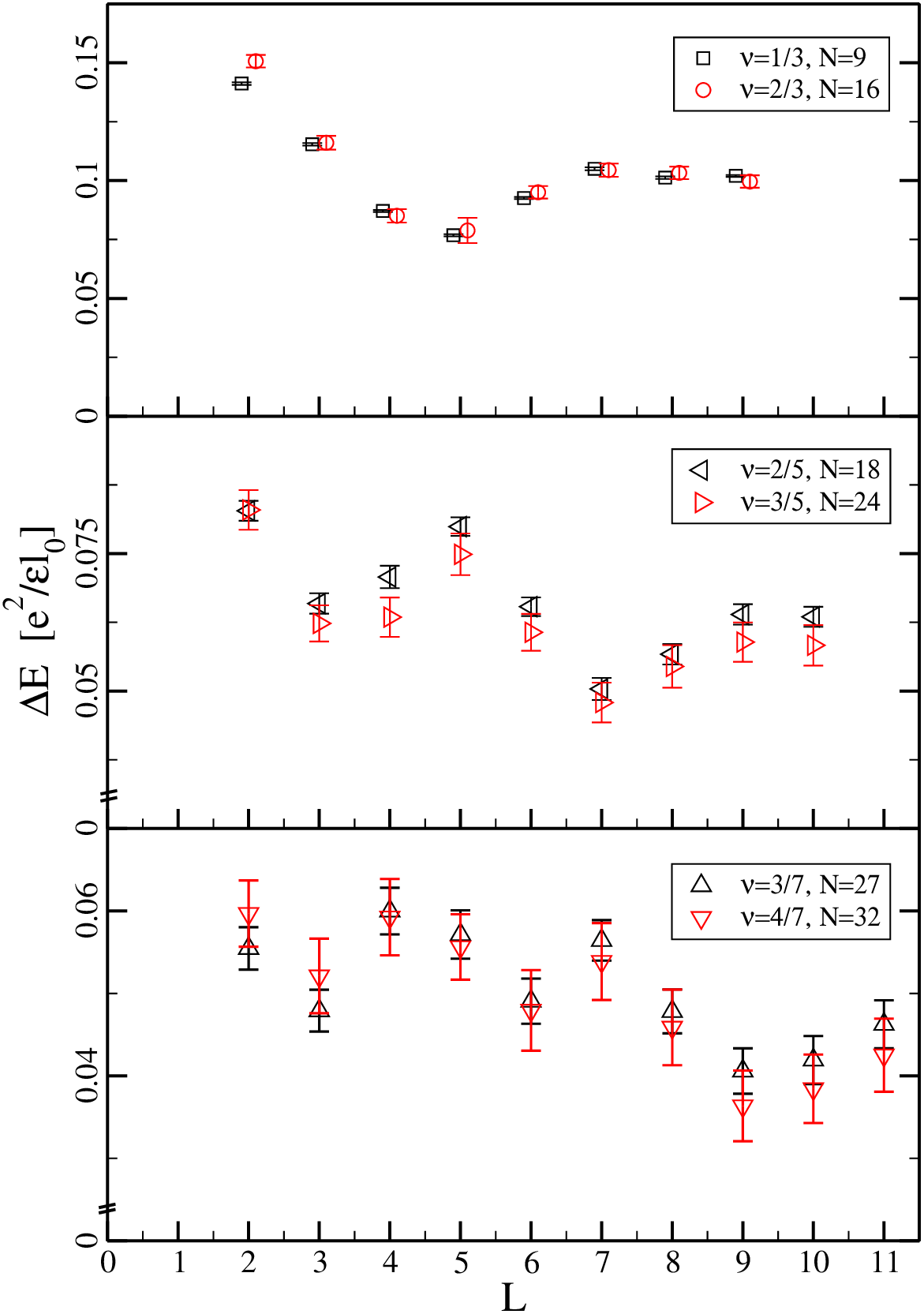}
\caption{ \label{fig:spectra_m1} Quantum Hall states that are
particle-hole conjugate of each other, are expected to have the
same excitation spectrum.  The low energy spectra shown here for
for $\nu < \frac{1}{2}$ have been calculated with the method of
Jain and Kamilla\cite{JainKamillaExciton}.  The spectra for the
particle-hole conjugate states with $\nu
> \frac{1}{2}$ are calculated using the new negative effective flux
trial wavefunctions described in this paper.  If the calculations
were exact (rather than just approximate) the corresponding
spectra of the particle-hole conjugate pairs would match exactly.
Although our calculations, being based on approximate trial
wavefunctions for ${}^2$CFs, are not exact, we still see
remarkably good agreement, suggesting that our new trial
wavefunctions are approximately as accurate as the previously
described trial wavefunctions for $\nu < \frac{1}{2}$. The spectra
shown here correspond (as marked) to particle hole conjugate pairs
at filling fractions, from top to bottom $1/3 \leftrightarrow
2/3$, $2/5 \leftrightarrow 3/5$ and $3/7 \leftrightarrow 4/7$
corresponding to flux $24, \,41 \text{ and } 58\,\phi_0$
respectively.  Note that our new trial wavefunctions show an
excellent reproduction of the increasingly nontrivial features as
$\nu=1/2$ is approached. In the top panel the data points have
been shifted slightly horizontally from integer angular momentum
values for better distinguishability.}
\end{figure}

The ground-state energies we obtain using negative $p$ trial
wavefunctions show an outstanding agreement with the values
obtained by particle hole conjugating positive $p$ trial
wavefunctions.  We observed slight differences on the fourth
significant digit, that show that the trial state with composite
fermions in positive effective flux is very slightly better than
the one with negative effective flux introduced here.

The excellent agreement of our negative $p$ wavefunctions with
particle-hole conjugation of positve $p$ wavefunctions extends to
the excited states, generated as outlined in section \ref{sec:jk}.
As examples, Figure \ref{fig:spectra_m1} shows excitation spectra
for $\nu=1/3$, $2/5$ and $3/7$ and their respective particle-hole
conjugate states.  We also give, in Table \ref{tab:results_m1}
values for magnetoroton gaps (which are the lowest energy neutral
excitations) as well as large $k$ gaps (which are presumably the
transport gap).   These results, along with the excellent results
for the ground-state energies confirm the validity of our approach
to calculating CF wavefunctions at negative effective flux, which
enables us to consider in the following section filling fractions
above 1/4 in the series $\nu=p/(4p+1)$, that were previously
inaccessible.

\begin{table*}[thpb]
\begin{tabular}{||c||c|c|c|D{.}{.}{5.8}|D{.}{.}{4.8}|D{.}{.}{4.8}||c|c|c|D{.}{.}{5.8}|D{.}{.}{4.8}|D{.}{.}{4.7}||}
\colrule \multicolumn{1}{||c||}{} & \multicolumn{6}{c||}{~~~~~~~~
Negative $p$ Trial Wavefunction~~~~~~~~~} &
\multicolumn{6}{c||}{~~~~~ Positive $p$ Trial Wavefunction~~~~~~}\\
\colrule
$\, N\,$&$p$&$N_\phi^\text{eff}$&$N_\phi$&\multicolumn{1}{c|}{$E_g$}&\multicolumn{1}{c|}{MR-gap}
&\multicolumn{1}{c||}{large $k$ gap } & $\, p\,$  &$N_\phi^\text{eff}$&$N_\phi$&
\multicolumn{1}{c|}{$E_g$}&\multicolumn{1}{c|}{MR-gap}&\multicolumn{1}{c||}{large $k$ gap}\\
\toprule
4&-1&-3&9&-0.47481(3)&0.0929(3)&0.1241(3)&1&3&15&-0.37706(1)&0.0169(1)&0.0305(1)\\
5&&-4&12&-0.45940(3)&0.0939(4)&0.1224(4)&&4&20&-0.36501(1)&0.0203(1)&0.0276(1)\\
6&&-5&15&-0.44996(3)&0.0837(4)&0.1183(4)&&5&25&-0.35800(1)&0.0132(1)&0.0251(1)\\
7&&-6&18&-0.44355(3)&0.0812(3)&0.1135(4)&&6&30&-0.35303(1)&0.0147(2)&0.0241(2)\\
8&&-7&21&-0.43895(3)&0.0830(4)&0.1110(5)&&7&35&-0.34955(1)&0.0151(2)&0.0243(2)\\
9&&-8&24&-0.43540(4)&0.0781(5)&0.1083(5)&&8&40&-0.34686(2)&0.0124(3)&0.0245(3)\\
20&&$-$&$-$&-&-&-&&19&95&-0.33577(2)&0.0109(5)&0.0242(5)\\
\cline{1-1} \cline{3-7} \cline{9-13}
-$\infty$&&$\infty$&$\infty$&-0.40999(4)&0.0668(5)&0.097(3)&&$\infty$&$\infty$&-0.32748(8)&0.009(2)&0.021(1)\\
\hline \hline
6&-2&-1&19&-0.40432(3)&0.0243(2)&0.0243(2)&2&1&21&-0.38699(1)&0.0182(2)&0.0182(2)\\
8&&-2&26&-0.39826(3)&0.0231(3)&0.0257(3)&&2&30&-0.37411(1)&0.0186(2)&0.0193(2)\\
10&&-3&33&-0.39484(3)&0.0190(3)&0.0283(3)&&3&39&-0.36726(1)&0.0142(4)&0.0220(3)\\
12&&-4&40&-0.39257(3)&0.0207(4)&0.0288(4)&&4&48&-0.36276(2)&0.0128(3)&0.0219(4)\\
14&&-5&47&-0.39092(3)&0.0179(5)&0.0275(5)&&5&57&-0.35969(2)&0.0125(4)&0.0208(5)\\
16&&-6&54&-0.38976(3)&0.0203(4)&0.0268(5)&&6&66&-0.35742(2)&0.0129(5)&0.0183(5)\\
18&&-7&61&-0.38885(3)&0.0176(8)&0.0235(9)&&7&75&-0.35569(2)&0.0108(6)&0.0179(4)\\
\cline{1-1} \cline{3-7} \cline{9-13}
$\infty$&&-$\infty$&$\infty$&-0.38163(3)&0.0151(1)&0.0276(2)&&$\infty$&$\infty$&-0.34277(5)&0.0077(13)&0.020(2)\\
\hline \hline
12&-3&-1&43&-0.38486(3)&0.0152(5)&0.0152(5)&3&1&45&-0.37306(2)&0.0132(4)&0.0132(4)\\
15&&-2&54&-0.38266(3)&0.0154(7)&0.0169(7)&&2&58&-0.36765(3)&0.0137(6)&0.0137(6)\\
18&&-3&65&-0.38126(3)&0.0140(8)&0.0190(8)&&3&71&-0.36427(3)&0.0119(7)&0.0171(6)\\
21&&-4&76&-0.38017(3)&0.0126(8)&0.0192(10)&&4&84&-0.36182(2)&0.0120(8)&0.0175(8)\\
24&&-5&87&-0.37940(3)&0.0126(10)&0.0196(9)&&5&97&-0.36006(4)&0.012(1)&0.017(1)\\
27&&-6&98&-0.37883(3)&0.0134(14)&0.0195(14)&&6&110&-0.35869(2)&0.010(1)&0.016(1)\\
\cline{1-1} \cline{3-7} \cline{9-13}
$\infty$&&-$\infty$&$\infty$&-0.37404(3)&0.010(1)&0.0238(8)&&$\infty$&$\infty$&-0.34842(4)&0.0088(12)&0.021(2)\\
\hline \hline
20&-4&-1&75&-0.37206(3)&0.0123(6)&0.0123(6)&4&1&77&-0.36788(3)&0.0109(7)&0.0109(7)\\
24&&-2&90&-0.37172(4)&0.010(1)&0.0117(10)&&2&94&-0.36491(3)&0.0118(9)&0.013(1)\\
28&&-3&105&-0.37165(4)&0.010(1)&0.0148(7)&&3&111&-0.36285(3)&0.011(1)&0.012(1)\\
32&&-4&120&-0.37142(3)&0.010(1)&0.013(1)&&4&128&-0.36133(3)&0.0093(9)&0.014(1)\\
36&&-5&135&-0.37138(2)&0.012(1)&0.016(1)&&5&145&-0.36021(3)&0.011(1)&0.014(2)\\
\cline{1-1} \cline{3-7} \cline{9-13}
$\infty$&&-$\infty$&$\infty$&-0.3705(1)&0.010(3)&0.019(3)&&$\infty$&$\infty$&-0.35129(5)&0.0087(19)&0.017(2)\\
\colrule
\end{tabular}
 \caption{Numerical results from Monte Carlo calculations
of ${}^4$CF wavefunctions for groundstate energies and gaps at
filling fraction $\nu=p/(4p+1)$, given in units of $e^2/\epsilon
\ell_0$. Alongside the newly calculated negative $p$ Quantum Hall
states of this series above $\nu=1/4$, we give the results for
states with the respective positive $p$ and equal particle number
for comparison.  We find that the similarity between the results
for $p$ and $-p$ increases with increasing $p$.  Two excitation
energies are given from the spectra of neutral excitations: the
magneto-roton gap as the lowest lying excitation and the large $k$
gap measured at the highest possible angular momentum.
Extrapolation to the thermodynamic limit has been performed using
the above given data by a simple linear regression over the
inverse particle number (see Figure \ref{fig:gap_thermo}). In the
case of groundstate energies, this extrapolation is based on
density corrected values\cite{Morf} ($E_g^\text{corr}=E_g\sqrt{\nu
N_\phi/N}$).} \label{tab:table_m2}
\end{table*}

\section{Results for $\nu=p/(4p+1)$}
\label{sec:cf4results}

\begin{figure}[pthb]
\includegraphics[width=.95\columnwidth]{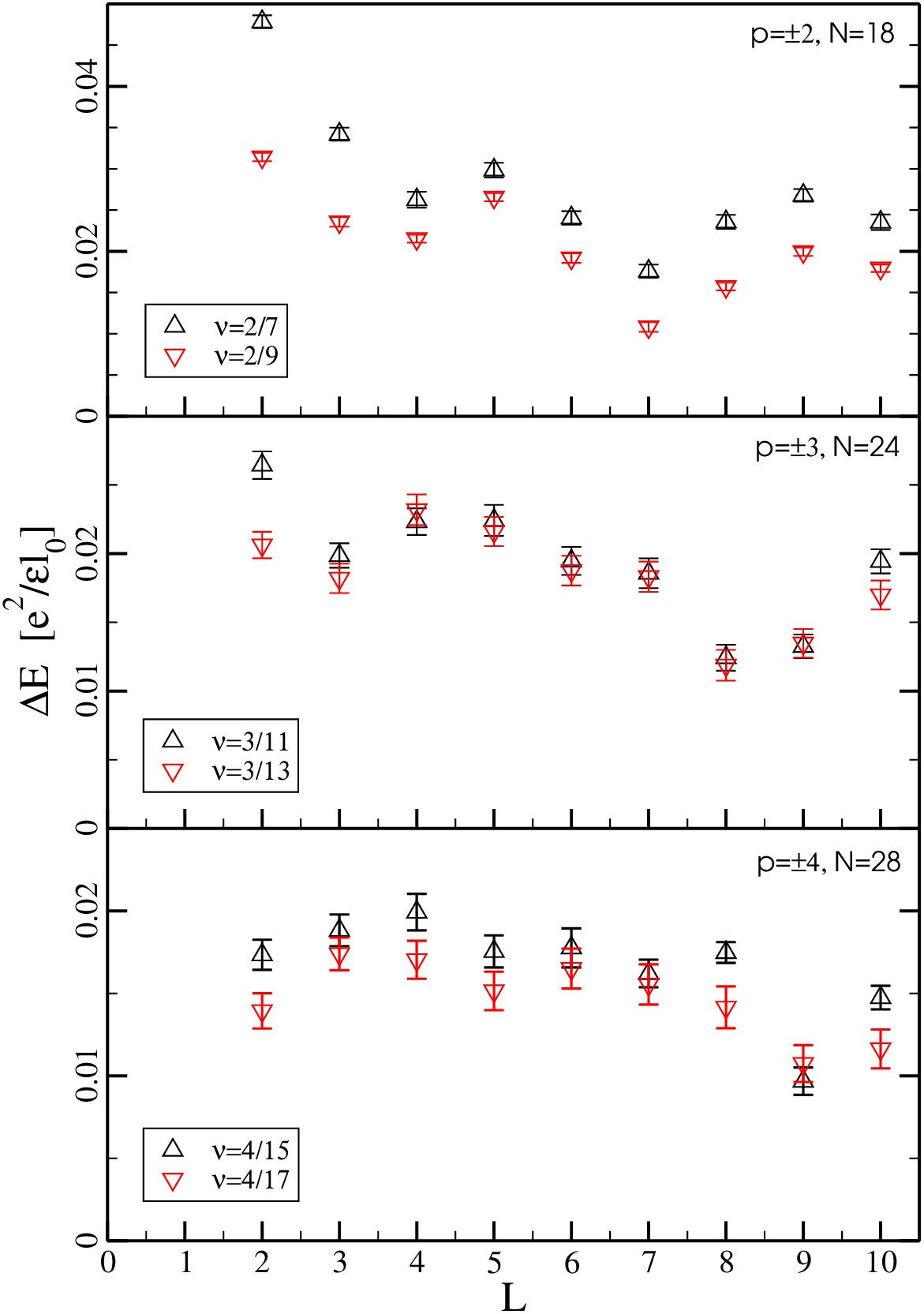}
\caption{ \label{fig:spectra_m2} The method developed here permits
us to calculate the dispersion curves for the low energy
excitations at filling fractions $p/(4p+1)$ for negative $p$, i.e.
above a quarter filling, that were previously inaccessible.  A
comparison to the spectra of states below and above $\nu=1/4$
shows, that the excitation spectra for $p$ and $-p$ show very
similar features and seem to become more similar as $p$ is
increased.      As above, the $p > 0$ spectra are calculated using
the method of Jain and Kamilla\cite{JainKamillaExciton} whereas
the $p < 0$ spectra are calculated using the method discussed in
the current paper.
}
\end{figure}

The ${}^4$CF series of composite fermion states, corresponding to
filling fractions around $\nu=1/4$, has been the subject of some
recent experimental work\cite{Pan}, yet the branch of filling
fractions above 1/4 has been mostly inaccessible to numerical
investigations. The numerical approach we took for examining these
states permits us to calculate the excitation spectra for systems
with a moderate number of effective magnetic flux quanta
$N_{\phi}^{\text{eff}} = 2q$ for the CF-system. For negative $p$,
the calculational complexity of the wavefunction increases with
$q$, consequently the achievable system size is reduced compared
to the systems with positive $p$. For this very reason,
extrapolation of the results to infinite system size in order to
obtain the gap in the thermodynamic limit, and thus the mass of
the CF, is difficult for these states. Nonetheless, the data
obtained so far reveals that the spectra of states with $p$ and
$-p$ at the same system size appear to be more and more similar as
$|p|$ increases and $\nu=1/4$ is approached. Table
\ref{tab:table_m2} summarizes the numerical results for the
groundstate energies and gaps of states around $\nu=1/4$, and
Figure \ref{fig:spectra_m2} shows several examples of excitation
spectra revealing the above mentioned similarity.

\begin{figure}[pbth]
 \includegraphics[width=1.00\columnwidth]{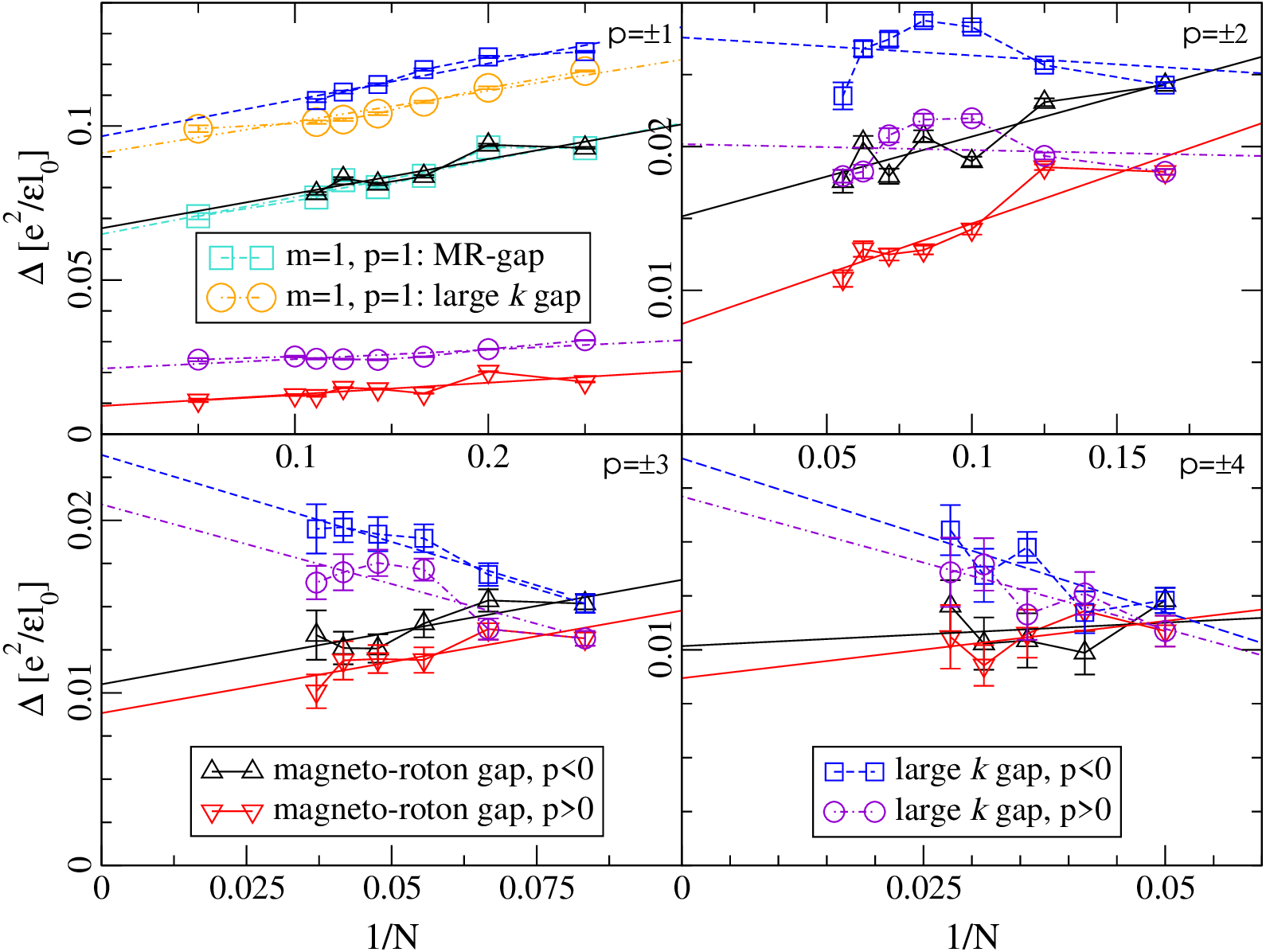}
 \caption{
 \label{fig:gap_thermo}
 This figure illustrates the extrapolation of the gaps to the thermodynamic
limit for different filling fractions in the series $p/(4p+1)$ by
means of a simple linear regression of the available data points
over the reciprocal particle number. States with $p$ and $-p$ are
displayed together. The state at $\nu=1/3$ can be obtained in two
different manners, i.e. as ${}^4$CF in negative flux or as
${}^2$CF in positive flux, which accounts for the two additional
sets of data in the upper left. The magnetoroton gaps of these
different $\nu=1/3$ states are close to indistinguishable, whereas
the large $k$ gap is slightly bigger for negative effective flux.
As discussed in the text, extrapolation is least certain for
$p=\pm 2$ due to finite size effects. }
\end{figure}

\begin{figure}[pthb]
\includegraphics[width=1.0\columnwidth]{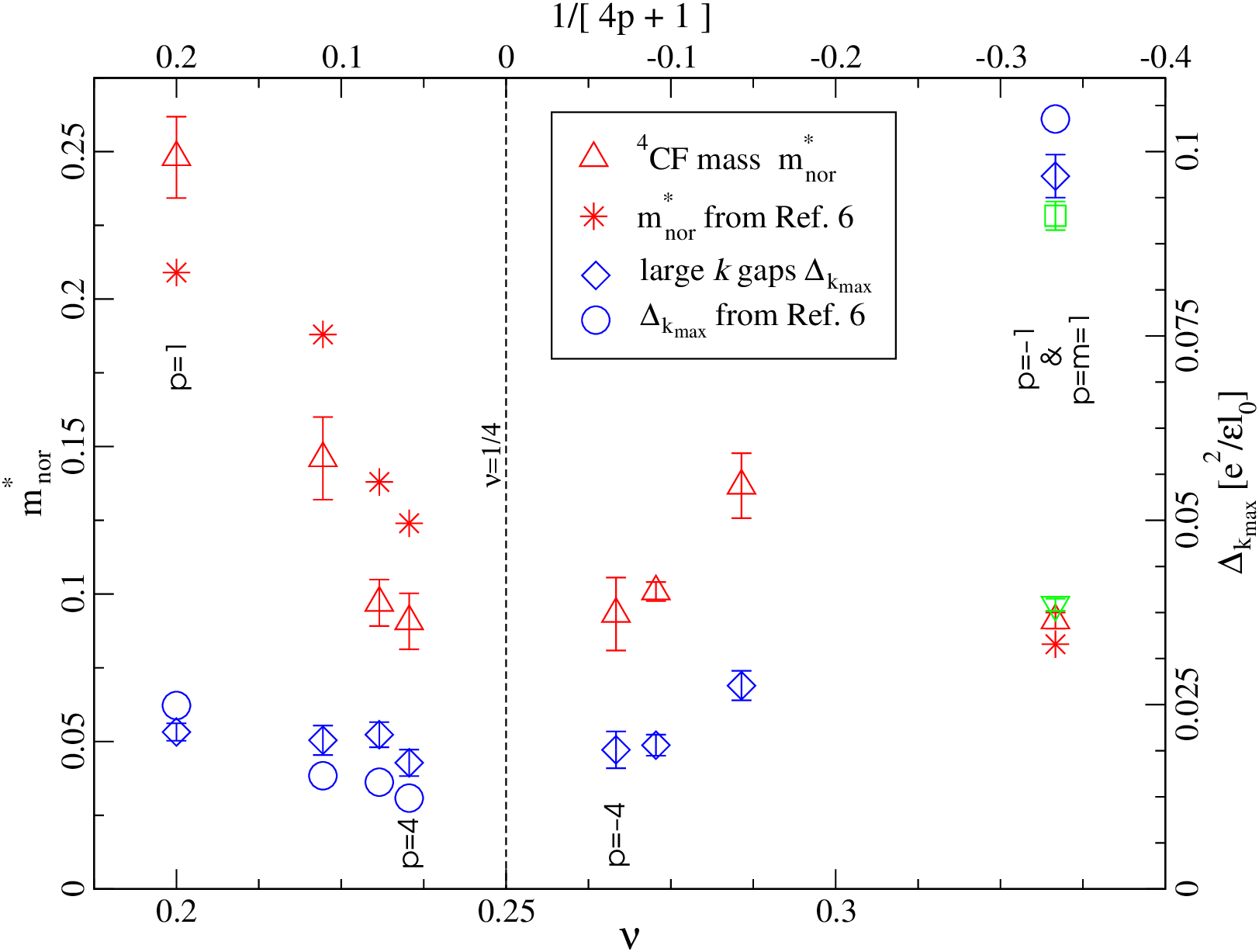}
\caption{ \label{fig:4CFmasses} Masses of the composite fermions
near $\nu=1/4$ and extra\-po\-lated large $k$ gaps from which they
were deduced at filling fractions of the series $\nu=p/(4p+1)$.
Data from Ref. \onlinecite{JainMass} is from larger system size
calculations and should be considered to be more accurate.
Comparing these more accurate results to our smaller system
calculations gives us an estimate of the finite size errors of our
results which presumably will hold even in the $p < 0$ case.
Except for the marked asymmetry between $\nu=1/3$ and $\nu=1/5$ we
do not see much sign of an asymmetry around $\nu=1/4$ which is
observed experimentally\cite{Pan}.  We also do not see signs of
the experimentally observed diverging effective mass as $\nu=1/4$
is approached.  Our values of the effective mass are somewhat less
than that obseved experimentally in general.  However, it is known
that finite well width corrections tend to increase the effective
masses\cite{JainMass}. Two additional data points are shown in
this figure (not mentioned in the legend) which give results
obtained by treating $\nu=1/3$ as a ${}^2$CF state.
}
\end{figure}

It is interesting to note that the $p \leftrightarrow -p$
similarity in spectra found here differs from the particle-hole
symmetry correspondence of $p$ with the $-p-1$ states in the
 $\nu=p/(2p+1)$ series shown above in section \ref{sec:phsym}.
One might naively expect that the particle-hole symmetry $p
\leftrightarrow -p-1$ for states in the ${}^2$CF series
$\nu=p/(2p+1)$ above survives the composite fermionization
attachment of two more Jastrow factors and results in an
approximate symmetry for $\nu=p/(4p+1)$ states. For example, we
could take the symmetry related $m=1$ pair ($p=-3, N=12$) and
($p=2, N=10$) at $N_{\phi} = 21$ on Table \ref{tab:results_m1} and
add two Jastrow factors so $N_{\phi} \rightarrow N_{\phi} +
2(N-1)$ resulting in an approximate symmetry relating ($p=-3,
N=12, N_{\phi} = 43$) to ($p=2, N=10, N_{\phi} = 39$) in Table
\ref{tab:table_m2}. However, when we examine the overall shape of
the resulting excitation spectra, we find very little relation
between the two spectra related by $p \leftrightarrow -p-1$ for
${}^4$CFs. In contrast we see in Figure \ref{fig:spectra_m2} that
the $p \leftrightarrow -p$ related states have quite similar
dispersions, and the similarity appears to increase with
increasing $|p|$.

We comment that another interesting test of our approach can be
obtained by constructing a trial wavefunction for $\nu=1/3$ as the
$p=-1$ member of the $\nu=p/(4p+1)$ series. We find that the
ground state energy of this wavefunction is, within numerical
precision, precisely the same as that of the Laughlin $\nu=1/3$
trial wavefunction, which leads us to believe that we have
precisely constructed that state.  Similarly, we can examine the
excitation spectrum of $\nu=1/3$ by using a ${}^2$CF wavefunction
of the series $p/(2p+1)$ with $p=1$ or by using  a ${}^4$CF
wavefunction of the series $p/(4p+1)$ with $p=-1$. We find that
the spectra obtained in these two approaches are quite similar to
each other (albeit not quite identical), which gives us still
further confidence in our approach. The values of the
magneto-roton gap appear to be almost exactly the same in both
cases. Yet, examining the large $k$ gap, the ${}^4$CF trial
wavefunction using negative flux ($p=-1$) yields a slightly larger
value, which decreases slightly more rapidly with the system size
though, and seems to extrapolate to almost the same value at
infinite $N$ (see Figure \ref{fig:gap_thermo}).

Figure \ref{fig:gap_thermo} shows the extrapolation to the
thermodynamic limit of the data given in Table \ref{tab:table_m2},
comparing gaps at $p$ with those of states at $-p$. A comparison
of the excitations at $\nu=1/3$ considered in the two different
manners discussed above is also displayed there. Further, this
figure gives us an idea of the quality of our extrapolation. Where
the extrapolation is not particularly smooth, we cannot claim to
deliver more than a rough result. The extrapolation of $p=\pm 2$
appears to be the most difficult, since it is not easy to
distinguish a clear linear dependence of the large $k$ gap as a
function of $N^{-1}$ for the initial data set. The reason for this
problem appears to be that the magnetoroton gap is located at a
large value of $L$, so that in the smaller systems, the
magnetoroton gap is located very close to the largest $k$
available, or even coincides with this point.
For the $p=2$ case, we can see the error in this extrapolation
clearly by comparing our extrapolated result to a result of a
similar calculation using larger system sizes from Ref.
\onlinecite{JainMass}.  This comparison is show in Figure
\ref{fig:4CFmasses} (see below).   We might guess that the error
in extrapolation for $p=-2$ is of similar magnitude.

 In order to obtain the  composite fermion effective
mass, we equate the activation gap $\Delta$ (determined from the
excitation energy at the maximum angular momentum, i.e. biggest
particle-hole separation) to the cyclotron energy of CFs in their
effective magnetic field:
\begin{equation}
 \label{eq:effective_mass1}
 \Delta_{\nu(m,p)}=\frac{\hbar e |B_{\subt{eff}}|}{m^\mystar\, c}=\frac{\hbar^2}
 {|2mp+1|\,m^\mystar\,\ell_0^2}
\end{equation}
Since the gap is measured in units of the Coulomb interaction, we
write $\Delta=(e^2/\epsilon\ell_0)\delta$. Further taking into
account $\epsilon_r=12.8$ for GaAs and the free electron mass as
our point of reference, we find the dimensionless
normalized\endnote{Note that Pan \emph{et. al.}\cite{Pan} use a a
different normalization for $m^\mystar$, which amounts to a slight
rescaling as follows:
$m^\mystar_{\subt{Pan}}=\sqrt{B_\nu/B_{1/4}}\,m^\mystar_{\subt{nor}}$.}
effective mass
$m^\mystar_\subt{nor}=m^\mystar/(m_e\sqrt{B_\nu[T]})$ to be given
by $m^\mystar_\subt{nor}=0.0264/(|2mp+1|\,\delta)$. The
${}^4$CF-masses we obtain are displayed in Figure
\ref{fig:4CFmasses} together with the large $k$ gaps from which
they are deduced.  In addition, for $\nu < 1/4$ (positive $p$) we
have shown data from Ref. \onlinecite{JainMass} where larger
systems were used than we have used here (When we also use larger
system sizes, our results agree very will with those of Ref.
\onlinecite{JainMass}). For $\nu=1/3$, as discussed above, we have
shown results that treat this either as a $p=-1$ point of the
$p/(4p+1)$ series or a $p=1$ point of the $p/(2p+1)$ series. For
the latter case, we have likewise included the result from the
above mentioned larger system calculation.

In Figure \ref{fig:4CFmasses}, we have intentionally displayed
data extrapolated from a restricted set of small systems sizes
with positive $p$ so as to match the same set of system sizes that
we study for negative $p$ where we cannot go to very large
systems. One can estimate the finite size error for our negative
$p$ calculations by examining the deviations between these
restricted extrapolations at positive $p$ compared to the larger
system calculations of Ref. \onlinecite{JainMass} (also displayed
in our figure).

The behavior of the negative $p$ fractions seems to roughly mirror
the behavior of the positive $p$ fractions.  As was seen
previously in Ref. \onlinecite{JainMass} for $p>0$ it is seen that the
effective mass increases with $|B_\text{eff}|$.  As $\nu=1/4$ is
approached from either side, it is not clear if the effective mass
will converge to a constant as would be predicted by theory (up to
logarithmic corrections\cite{HLR}).  In the experiments of Ref.
\onlinecite{Pan}, a striking asymmetry of the effective mass beween the
high field and low field sides of $\nu=1/4$ has been observed.   While we cannot
rule out some asymmetry from our data, we certainly cannot claim
to see the extremely strong differences that are observed
experimentally.   This, however, is not surprising.
Experimentally, the asymmetry is attributed to the proximity of a
Wigner crystal state\cite{Pan}.  Since we are using a trial
wavefunction approach, we should not see the effects of any
imminent phase transition.

Perhaps the most interesting data point in Figure
\ref{fig:4CFmasses} is the one at $\nu=1/3$.  Whether we treat
this point as the $p=-1$ member of the series $p/(4p+1)$ or the $p=1$
member of the series $p/(2p+1)$, we find almost identical results
of a very large gap, which establishes a continuity between the CF
masses around $\nu=1/2$ and $\nu=1/4$. Furthermore, we note that
this point is quite asymmetric with its reflection at $\nu=1/5$.
Certainly the hypothesis of constant effective composite fermion
mass does not extend all the way from $\nu=1/4$ out to both
$\nu=1/5$ and $\nu=1/3$. Below quarter filling, one observes a
continuous increase of this mass. A similar trend appears  at
small values of the effective magnetic field above this point, but
then the mass drops down again at $\nu=1/3$, the final point of
this series.

Generally, our values for the effective mass seem to be lower than
those measured in the experiments of Ref. \onlinecite{Pan} by a
factor of roughly $2.5$.  This error is rather expected, since
similar discrepancies have been observed in previous studies based
on the composite fermion picture\cite{JainMass}. It is known,
however, that taking into account the finite width of the 2D
electron gas changes the interaction so as to increase the
effective mass\cite{JainMass}.

We note that we are not able to find any evidence of the
divergence of $m^\mystar$ as we approach $\nu=1/4$ from either
side, which is observed experimentally in Ref. \onlinecite{Pan}.
This is not surprising for several reasons. First of all, the
experiment only sees strong divergences extremely close to
$\nu=1/4$ -- which we cannot access numerically.  Furthermore, one
might suspect that the disorder might be the source of the
divergences in the measurements used by Pan. More importantly,
however, even if there were genuine infra-red
divergences\cite{HLR} of the effective mass as $\nu=1/4$ is
approached, one would not necessarily expect such divergences to
be properly represented in a trial wavefunction approach.

We emphasize that the most important achievement of this paper is
not any particular numerical result.  If genuine numbers were
desired for comparison to experiment, we would want to use a more
realistic interaction, accounting for finite well
width\cite{JainMass}, as well as perhaps Landau-level mixing, and
we would want to use a more powerful computer to analyze ever
larger systems.  Instead we would like to emphasize in this paper
that we have clearly demonstrated that we can extend the approach
of Jain and Kamilla\cite{JainKamilla} to treat negative effective
magnetic field, and we can study these negative $p$ composite
fermion wavefunctions for reasonably large systems, which has not
been done before.   In this paper we have tested this method by
using particle-hole symmetry for the ${}^2$CF series and we have
found our method to be quite accurate.  We have then applied this
method to ${}^4$CFs to study, for the first time, large systems
for filling fractions $p/(4p+1)$ with negative $p$.   Our main
physical result is that the effective mass appears to be roughly
symmetric around and close to $\nu=1/4$ although larger system
calculations would be desirable.

The authors acknowledge helpful conversations with E. H. Rezayi.
G.M. acknowledges support from the French Ministry of Science, and
thanks both the Ecole Doctorale de Paris and Lucent Technologies
for their support that made participation in this research project
possible.

\appendix

\section{CF wavefunctions with negative effective flux}
\label{app:jk}

The starting point for the composite fermion trial
wavefunctions\cite{JainKamilla}
are the single particle eigenfunctions of the
quantum mechanical problem of a particle in a magnetic monopole
field on a sphere.   (We use spherical coordinates with the
azimuth $\theta$ ranging from $0$ to $\pi$, and $\phi$ the
longitude ranging from $0$ to $2 \pi$). The monopole harmonics are
given by\cite{WuYang}
$$
Y^q_{n,m}(\Omega) = 2^m M_{q,n,m}
(1-x)^{\alpha/2}(1+x)^{\beta/2}P_g^{\alpha,\beta}(x)e^{im\phi}
$$
with  $\alpha=-q-m$, $\beta=q-m$, $g=|q|+n+m$, $x=\cos\theta$,
$$
M_{q,n,m}=\sqrt{\frac{2|q|+2n+1}{4\pi}\frac{(|q|+n-m)!(|q|+n+m)!}{n!(2|q|+n)!}}
$$
and $P_g^{\alpha,\beta}(x)$ are the Jacobi polynomials. This
monopole harmonic represents an eigenstate of a particle on a
sphere in a radial magnetic field with $2q$ flux quanta
penetrating the sphere, where a positive sign refers to outwards
pointing flux.  Here, the angular momentum of the eigenstate is
$l=|q|+n$ and the  $z$ component of the angular momentum $m\in
\{-l,\ldots ,l\}$.  Further, $n$ is the LL index $n =0,1,\ldots$.
The above expression assumes the Haldane gauge\cite{Haldane},
where the singularities of the vector potential are chosen
to be located
symmetrically on both north- and southpole of the sphere.  Here, we
focus on the case where $q<0$, since the $q> 0$ case has
already been discussed in detail in Ref. \onlinecite{JainKamilla}.
For the rest of this paragraph, we thus assume $q<0$.
Expanding the Jacobi polynomials in terms of the spinor
coordinates $u=\cos(\theta/2)e^{-i\phi/2} $ and
$v=\sin(\theta/2)e^{i\phi/2}$, one obtains
\begin{eqnarray}
\label{Y_in_u_v}
& & Y^{q<0}_{n,m}(\Omega) = (-1)^n M_{q,n,m}\,(u^\mystar)^{-q+m}(v^\mystar)^{-q-m} \times\nonumber\\
& &  \sum_{s=0}^n(-1)^s
\binom{n}{s}\binom{2|q|+n}{|q|+m+s}(u^\mystar u)^s(v^\mystar v
)^{n-s}
\end{eqnarray}
This expression can equally well be obtained from the relation for
complex conjugation of the monopole harmonics\cite{WuYang2}, if one corrects
Kamilla's formula by replacing $q$ by $|q|$ in the appropriate places.
We now use the $Y^{q}_{n,m}$ as single particle wavefunctions
(written as $\psi_i$ in the main text) and composite fermionize by
attaching Jastrow factors. As discussed in the main text, we can
construct the many particle composite fermion trial wavefunctions
by bringing the Jastrow factors inside the Slater determinant of
single particle states, which may then be projected individually.
Note, that in the spherical geometry, the Jastrow factor becomes
\begin{equation}
   J_j = \prod_{k \neq j} (u_k v_j - u_j v_k)
\end{equation}
The details of the required projection
$\mathcal{P}(Y^q_{n,m}(u_i,v_i)J_i^p(u_1,v_1,\ldots,u_N,v_N))$ are
discussed next.

First, we remark that the Jastrow factor\footnote{in this
appendix, we note the number of flux pairs attached to each CF as
$p$, according to Kamilla's notation, whereas $m$, that had been
used for this purpose in section \ref{sec:jk} to comply with the
common notation for the filling fractions of Jain's series, is
already used for the eigenvalue of $L_z$ here.} $J_i^p$ is a LLL
function, with $q'=p(N-1)$ zeros in $u_i$, i.e. it is a LLL
wavefunction for flux $q'>0$. Since $N$ is in general a big
number, we have $q'\gg |q|$. The resulting wavefunction, in turn,
has to be a valid wavefunction for a total number of flux $Q = q +
q'>0$. Secondly, since projection is a linear operation, we may
consider the action of projection on each of the basis states
$Y^{q'}_{0,m'}$ separately, by expanding $J_i^p$ in this basis. In
general multiplication by a basis state $Y_{n,m}^q$ followed by
projection can be described as a linear operator called hereafter
$\mathfrak{Y}^{q'}_{q,n,m}$.
\begin{align}
\label{Projection_of_Basis_states}
\mathcal{P}Y^q_{n,m}Y^{q'}_{0,m'}=
\mathfrak{Y}^{q'}_{q,n,m}Y^{q'}_{0,m'}
\end{align}
Since we know the entire basis of the subspace that we project upon,
namely the LLL for flux $Q$ with states $|M\rangle$, and $|M|\leq Q$,
the projection operator is $\sum_M |M\rangle\langle M|$. We now show,
how this leads to an expression for $\mathfrak{Y}^{q'}_{q,n,m}$ as a
differential operator in the coordinate representation, in which
(\ref {Projection_of_Basis_states}) becomes
\begin{align}
\label{Projection_in_coordinate_rep}
\sum_{M=-Q}^Q Y^Q_{0,M}(\Omega) \int d\Omega'\,Y^{Q\phantom{k}\mystar}_{0,M}(\Omega') Y^q_{n,m}(\Omega') Y^{q'}_{0,m'}(\Omega')\nonumber\\
=\mathfrak{Y}^{q'}_{q,n,m}Y^{q'}_{0,m'}(\Omega)
\end{align}
Integration over the longitudinal angle $\phi$ singles out one
nonzero scalar product for $M=m+m'$, and the one remaining
integral over the azimuthal angle $\theta$ yields a well known
binomial coefficient. Simplifying the normalization factors of
$Y^{q'}_{0,m'}$ on both sides, we have:
\begin{align}
&(-1)^n M_{q,n,m} \sum_s (-1)^s\binom{n}{s}\binom{2|q|+n}{|q|+m+s} \nonumber\\
\times&(N_{Q,0,M})^2 4 \pi\frac{(q'-m'+s)!(q'+m'+n-s)!}{(2q'+n+1)!} \nonumber\\
\times& u^{q-m+q'-m'} v^{q+m+q'+m'}
=\mathfrak{Y}^{q'}_{q,n,m} u^{q'-m'} v^{q'+m'}
\end{align}
Using the explicit form of the normalization
\begin{equation}
( N_{Q,0,M})^2 = \frac{(2Q+1)!}{4\pi (Q+M)!(Q-M)!}
\end{equation}
and remarking that the fractions of factorials that
are left in this expression equal those that appear
by the multiple derivation of a monomial $u^k$
\begin{align}
\left(\frac{\partial}{\partial u}\right)^{s-q+m}u^{q'+m'+s}
=\frac{(s+q'+m')!}{(Q+M)!}u^{Q+M}
\end{align}
we may deduce $\mathfrak{Y}^{q'}_{q,n,m}$ by comparison of
both sides:
\begin{align}
\label{eq:ProjOp}
\mathfrak{Y}^{q'}_{q,n,m}=&\frac{(2Q+1)!}{(2q'+n+1)!}(-1)^n M_{q,n,m}\times\nonumber\\
&\sum_{s=0}^{n}(-1)^s\binom{n}{s}\binom{2|q|+n}{|q|+m+s} \times\nonumber\\
&\left(\frac{\partial}{\partial
u}\right)^{|q|+m+s}u^s\left(\frac{\partial}{\partial
v}\right)^{|q|-m+n-s}v^{n-s}
\end{align}
Let us remark that this result reproduces the known result, that
the projection on the LLL is achieved by performing the habitual
procedure of moving all $u^\mystar$'s and $v^\mystar$'s to the far
left, and replacing them with derivatives according to
\begin{align}
u^\mystar \rightarrow \frac{\partial}{\partial u}, \text{
and\quad} v^\mystar \rightarrow \frac{\partial}{\partial v}
\end{align}
Nevertheless, performing this explicit projection gives us a supplementary
information in the form of a weight factor $\frac{(2Q+1)!}{(2q'+n+1)!}$
for the different Landau levels before projection, which of course
does not matter for the problems discussed here, but may play a
role in other cases\cite{SimonMoller}.

Practically, we would like to obtain a form of (\ref{eq:ProjOp}) with the
derivatives moved to the extreme right, which may be calculated using a straightforward
application of the Leibniz rule for multiple derivatives in both $u$ and $v$:
\begin{equation}
\label{eq:Leibnitz}
\left( \frac{\partial}{\partial v}\right)^\beta v^\gamma =\sum_{\alpha=0}^\beta \frac{\beta !}{\alpha !} \binom{\gamma}{\beta-\alpha}v^{\gamma-\beta-\alpha}\left( \frac{\partial}{\partial v}\right)^\alpha
\end{equation}
This yields a triple sum with the inner summation ranges being
dependant on the outer summation index $s$. One finds that the
summation ranges can be made independent of $s$ since the summand
becomes zero outside of the given intervals. As such the sum over
$s$ may be evaluated using
\begin{equation}
\sum_{s=0}^{n}(-1)^s \binom{n-\alpha-\alpha'}{s-\alpha} = (-1)^\alpha \delta_{n,\alpha+\alpha'}
\end{equation}
Since the result yields a Kronecker delta, one of the remaining
sums becomes trivial, and after shifting the remaining summation
index, the final result is revealed to be exactly like
(\ref{eq:ProjOp}), but with all derivatives placed at the very
right.

The projected composite fermion wavefunction is nothing but this
operator applied to the single particle Jastrow factor:
\begin{align}
Y^{q\phantom{,m}CF}_{n,m}(\Omega_i)=\mathfrak{Y}^{q'}_{q,n,m}J_i^p
\end{align}

In order to perform numerical calculations with this wavefunction,
we need to evaluate the derivatives explicitly. One may use Jain
and Kamilla's approach\cite{JainKamilla}, to commute the
derivatives through the Jastrow factors as
\begin{align}
&\left(\frac{\partial}{\partial
u_i}\right)^s\left(\frac{\partial}{\partial v_i}\right)^t
J_i^p=J_i^p[\hat U_i^s \hat V_i^t 1] \intertext{with} &\hat U_i =
J_i^{-p} \frac{\partial}{\partial u_i}J_i^p,\text{ and }\,\hat V_i
= J_i^{-p} \frac{\partial}{\partial v_i}J_i^p
\end{align}
Constructing a many-particle wavefunction out of these projected
CF wavefunctions in the form of a Slater determinant, one may
factor out the Jastrow factors again, and thus obtains a form
which resembles single particle wavefunctions, on a basis of
projected states $\tilde Y^{q}_{n,m}$ with:
\begin{align}
\tilde Y^{q}_{n,m}(\Omega_i)=&\frac{(2Q+1)!}{(2q'+n+1)!}(-1)^{n} M_{q,n,m}\times\nonumber\\
&\sum_{s=0}^{n}(-1)^s\binom{n}{s}\binom{2|q|+n}{|q|+m+s} \times\nonumber\\
&u_i^s v_i^{n-s} [ \hat U_i^{|q|+m+s} \hat V_i^{|q|-m+n-s} 1 ]
\end{align}
Of course, this only appears to be a one particle wavefunction,
since there is an implicit dependence of the positions of all other
electrons in the system hidden in the operators $\hat U_i$ and
$\hat V_i$. The complexity of this expression increases with the
total number of derivatives per term, given by $N_{\partial}^{q<0}=2|q|+n$ for
negative $q$ compared to $N_{\partial}^{q>0}=n$ for positive
effective flux.

\section{Particle-hole Conjugation}
\label{app:phconj}

In the lowest Landau level, on the sphere, there are $N_\phi+1$
single particle eigenstates where $N_\phi$ is the flux through the
sphere.  We label these eigenstates by the $z$-component of their
angular momentum $m$.

For a  two-body interaction we can write the Hamiltonian as
\begin{equation}
 H=\sum _{m_1,m_2,m_3,m_4} V_{m_1,m_2,m_3,m_4} c^\dagger_{m_1}c_{m_2}^\dagger c_{m_3}c_{m_4}
\end{equation}
As usual, the normal ordering of the operators accounts for the
uniform positive background. As usual, the fermion operators have
anticommutation relations $ \{c_i^\dagger,c_j\}=\delta_{i,j} $.
The interaction matrix $V$ must have the following symmetries
\begin{eqnarray} \nonumber
V_{m_1,m_2,m_3,m_4} &=& -  V_{m_2,m_1,m_3,m_4} = -
V_{m_1,m_2,m_4,m_3}
\\ &=&  V_{m_4,m_3,m_2,m_1}^*
\end{eqnarray}
Furthermore, for any rotationally (translationally) invariant
interaction, we must have angular momentum conservation, which
implies that the matrix element is zero unless
\begin{equation}
\label{eq:angcons}
   m_1 + m_2 = m_3 + m_4
\end{equation}

We define the vacuum state $|0_e\rangle$, to be the state which
contains no electrons at all.  The filled Landau level, which we
can also think of as the vacuum for holes is written as
$|0_h\rangle = |\text{Filled}_e \rangle = \prod_{m }c_m^\dagger
|0_e\rangle$.  It is convenient to introduce creation and
annihilation operators $d,d^\dagger$ for holes, given by
$d_i=c_i^\dagger\, ,\;d_i^\dagger=c_i$ which also obey the usual
anticommutations $\{d_i^\dagger,d_j\}=\delta_{i,j}$. We now
rewrite the Hamiltonian in terms of these hole operators. Using
the commutation relations as well as above described symmetries of
$V$ we obtain  \begin{eqnarray} \nonumber H&=&
\sum_{m_1,m_2,m_3,m_4} V_{m_1,m_2,m_3,m_4} d_{m_1} d_{m_2}
d^{\dagger}_{m_3}
d^{\dagger}_{m_4}\\
&= & 2 \sum_{m} U_{m} (1 - 2\, d^\dagger_{m} d_{m}) + H_d
\end{eqnarray}
where
\begin{equation}
\label{eq:Udef}
   U_{m} = \sum_{m_2} V_{m, m_2, m_2, m}
\end{equation}
and
\begin{equation}
H_d = \sum_{m_1,m_2,m_3,m_4} V_{m_1,m_2,m_3,m_4}^*
d^\dagger_{m_1}d^\dagger_{m_2}d_{m_3}d_{m_4} \label{eq:Hddef}
\end{equation}

We show below that for any rotationally invariant interaction,
$U_m$ is actually independent of $m$.  Furthermore, it is very
easy to show that the energy of the entirely filled Landau level
is given by
\begin{equation}
 E_{\subt{filled}} = 2 \sum _{m_1,m_2} V_{m_1,m_2,m_2,m_1} = 2
 \sum_{m} U_m
\end{equation}
Thus we have
\begin{equation}
H=\Bigl( 1-\frac{2 N_h}{N_\phi+1} \Bigr) E_{\subt{filled}} + H_d
\end{equation}
and $N_h$ is the number of holes (i.e., the eigenvalue of $\sum_m
d^\dagger_m d_m$).

A general state containing $N$ electrons in the LLL can be written
as:
\begin{equation}
\label{eq:electron_state} |\Psi\rangle = \sum_{\{m_i\}}
a_{m_1,...,m_{N}} c^\dagger_{m_1}\cdots c^\dagger_{m_{N}}
|0_e\rangle
\end{equation}
To particle-hole conjugate this state we construct
\begin{equation}
\label{eq:hole_state} |\phc \Psi\rangle = \sum_{\{m_i\}}
a_{m_1,...,m_{N}}^* d^\dagger_{m_1}\cdots d^\dagger_{m_{N}}
|0_h\rangle
\end{equation}
which is now a state containing $N$ holes, or $N_\phi  + 1 - N$
electrons. Note that both of these states ``live" in the same
lowest Landau level which has  $N_\phi + 1$ single particle
eigenstates (indexed by $m$).

Now, if $\Psi$ is an eigenstate of $H$ with eigenvalue $E_{\Psi}$
(assumed to be real), then, since $H_d$ has precisely the same
structure as $H$ we see that $\phc \Psi$ is an eigenstate of
$H_d$ with the same eigenvalue.  Thus, we obtain the particle hole
conjugation relation
\begin{equation}
E_{\Psi}=\Bigl( 1-\frac{2 N_h}{N_\phi+1} \Bigr) E_{\subt{filled}} +
E_{\phc \Psi}.
\end{equation}

{\it {Lemma: $U_m$ is independent of $m$}:}

If an interaction is
rotationally invariant, we can choose any rotation $R$ and write
\begin{widetext}
\begin{equation}
   V_{m_1,m_2,m_3,m_4} =  \sum_{m_1', m_2', m_3', m_4'} D_{m_1,m_1'}^l(R)D_{m_2,m_2'}^l(R)
   \,\, V_{m_1',m_2',m_3',m_4'}
 \,\,  [D_{m_3,m_3'}^l(R)]^*  [D_{m_4,m_4'}^l(R)]^*
\end{equation}
\end{widetext}
where the $D$'s are rotation matrices as in Ref.
\onlinecite{Edmonds} and $l = 2 N_{\phi}$.  Setting $m_2 = m_3$,
and $m_1 = m_4 = m$ and summing over $m_2$ as prescribed in Eq.
\ref{eq:Udef} we obtain
\begin{equation}
   U_m = \sum_{m_1', m_2'} D_{m,m_1'}^l(R) \,\, V_{m_1',m_2',m_2',m_1'}
 \,\,  [D_{m,m_1'}^l(R)]^*
 \label{eq:Ures}
\end{equation}
where we have used the orthogonality\cite{Edmonds}
\begin{equation}
\sum_{m_2}     D_{m_2,m_2'}(R) [ D_{m_2,m_3'}(R)]^* =
\delta_{m_2',m_3'}
\end{equation}
as well Eq. \ref{eq:angcons}.  Since Eq. \ref{eq:Ures} must be
true for any rotation, we can integrate over all rotations and
use\cite{Edmonds}
\begin{equation}
   \int dR\, D_{m,m_1'}^l(R)
  [D_{m,m_1'}^l(R)]^* = \mbox{constant}
\end{equation}
independent of $m$, which shows that $U_m$ is independent of $m$.

\end{document}